# Beam Dynamics Design and Electromagnetic Analysis of 3 MeV RFQ for TAC Proton Linac

A. Caliskan[1], *H.F. Kisoglu[2,3], M. Yilmaz[3]

[1]*Department of Physics Engineering, Gumushane University, Gumushane, Turkey*
[2]*Department of Physics, Aksaray University, Aksaray, Turkey*
[3]*Physics Department, Gazi University, Ankara, Turkey*

**Abstract**

A beam dynamics design of 352.2 MHz Radio-Frequency Quadrupole (RFQ) of Turkish Accelerator Center (TAC) project which accelerates continuous wave (CW) proton beam with 30 mA current from 50 keV to 3 MeV kinetic energy has been performed in this study. Also, it includes error analysis of the RFQ in which some fluctuations have been introduced to input beam parameters to see how the output beam parameters are affected, two-dimensional (2-D) and three-dimensional (3-D) electromagnetic structural design of the RFQ to obtain optimum cavity paramaters that agree with the ones of the beam dynamics. The beam dynamics and error analysis of the RFQ have been done by using LIDOS.RFQ. Electromagnetic design parameters have been obtained by using SUPERFISH for 2-D cavity geometry and CST Microwave Studio for 3-D cavity geometry.

---

* Corresponding author. E-mail:hasanfatihk@aksaray.edu.tr

## 1. Introduction

The Turkish Accelerator Center (TAC) project [1] was approved by Turkish State Planning Organization (DPT) in 2006. The components of the project, which are an Infrared Free Electron Laser (IR-FEL) & Bremstrahlung facility, a particle factory, a third generation Synchrotron Radiation (SR) facility, a Self-Amplified Spontaneous Emission (SASE) mode Free Electron Laser (FEL) and a GeV scale linear proton accelerator (proton linac) facility, are being developed by more than 10 Turkish universities collaboration [2].

The envisaged proton linac will accelerate the proton beam up to 2 GeV and serve as a source that is effective use of many industrial, technical and health service areas to the users. It will also provide opportunity of research in nuclear science and high energy physics. The primary objective is usage of this linac in energy generation based on Accelerator Driven Systems (ADS) technology in the view of thorium reserves of Turkey [3].

The proposed linac will consist of a low-energy section of ~3 MeV, a medium-energy section of ~250 MeV and a high-energy section of ~2 GeV with superconducting cavities (Figure 1). The low-energy section, front-end of the linac, will be composed of a microwave-off resonance type ion source, a low energy beam transport (LEBT) line that transports and matches the beam from the source with the RFQ, and a radio-frequency quadrupole (RFQ) which is "sine qua non" for today's heavy ion linacs [4].

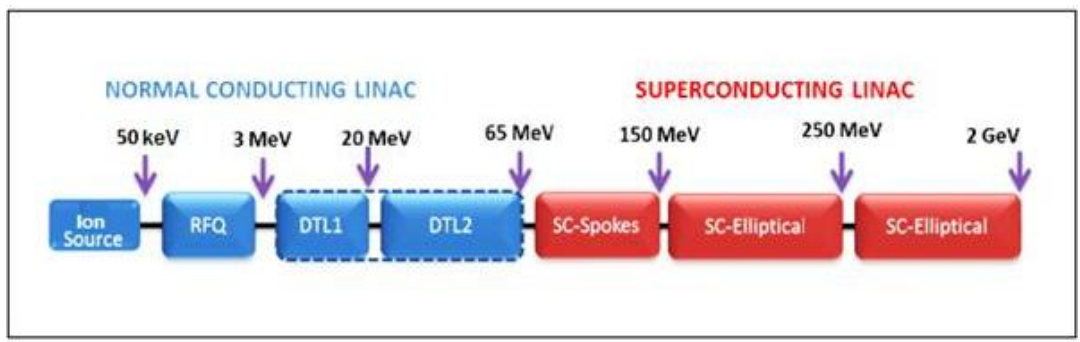

**Figure 1**. The block diagram of TAC proton linac

In this paper, we focus on beam dynamics based design as well as error analysis and electromagnetic structural design of RFQ. The design specifications of the RFQ are given in Table 1.

Table 1. Design requirements of the RFQ

| Parameters | Value | Unit |
|---|---|---|
| RFQ type | 4-vane | |
| Frequency | 352.2 | MHz |
| Duty cycle | 100% (cw) | |
| Particle | Proton | |
| Beam current | 30 | mA |
| Input energy | 50 | keV |
| Output energy | 3 | MeV |

We have used an input beam with a current of 30 mA for beam dynamics of the RFQ in accordance with the latest feasibility studies. A four-vane type RFQ has been chosen pursuant to 352.2 MHz radio-frequency. Also, cw beam has been envisaged to meet a need of the various applications. So, the power consumptions must be tackled cautiously. Beam dynamics and error anaylsis of the RFQ have been simulated using LIDOS.RFQ [5] in this paper. Also, electromagnetic structural design parameters were obtained by using SUPERFISH [6] for 2-D cavity geometry and CST Microwave Studio [7] for 3-D cavity geometry.

## 2. Beam Dynamics Design of the RFQ

The design studies of the RFQ based on beam dynamics simulation are performed optimizing beam dynamics parameters in compliance with given conditions such as operating RF frequency, intervane voltage, input beam current, kinetic energy, emittance, etc. The desired energy and beam current at the exit of the RFQ must be achieved using the parameters obtained from the beam dynamics design. Minimum emittance growth, compactness of the RFQ and maximum beam transmission are another requirements for the beam dynamics design of an RFQ.

Our RFQ specifications are given in Table 1. The emittance of the input beam has been chosen to be 0.20 $\pi\cdot$mm$\cdot$mrad (normalized, rms). The input energy should be selected as low as space-charge forces permit to get more compact structure. The initial particle distribution has been chosen as 4-D Uniform with 10000 particles.

Some beam dynamics parameters, which are modulation parameter ($m$) and synchronous phase ($\Phi_s$), have been tuned using LIDOS.RFQ software taking into account space-charge effects. Evolution of these two parameters and other beam dynamics parameters such as minimum aperture (*a*), acceleration effiency (*A*), kinetic energy (*W*) and intervane voltage ($U_0$) in consequence of the optimization are shown in Figure 2.

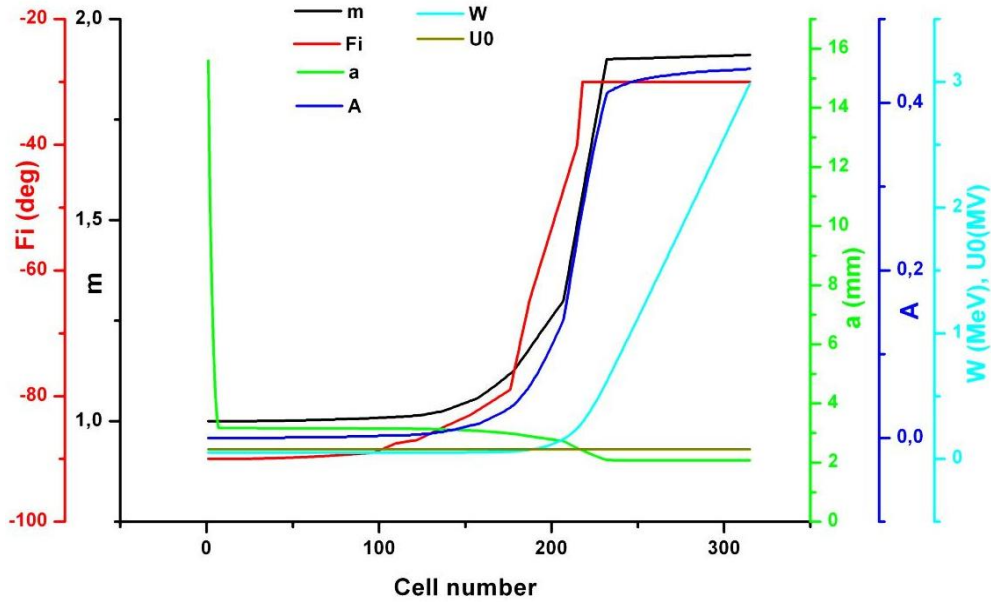

**Figure 2.** Evolution of some beam dynamics parameters along the RFQ in consequence of the optimization

RFQ structure is divided into four sections in conventional design methods. One of these is "Radial Matching Section (RMS)" that adapts cw beam to time varying structure of RFQ. We have reserved 7 cells, with 0.44 cm length for each, for RMS in our design. The structure consists of 316 cells which make a length of 3.45 m in total. As it is seen from the Figure 2, minimum aperture, *a*, from the *z*-beam axis goes down from a maximum value of 15.6 mm to the average bore radius, $r_0$, of 3.17 mm in the RMS. The *m* and $\Phi_s$ have values of 1, which means there is no modulation, and -90°, respectively, in this section. This means that there is no acceleration while focusing is maximum in this section. Thus, the beam is not formed into bunches and the bucket has the maximum length [8].

"Shaper" section comes after the RMS downstream of the beam. This section regulates the parameters as required by "Gentle Buncher (GB)" section which follows the shaper. 116 cells in the shaper give the beam a small acceleration since *m* rises up to 1.015. There is still focusing although not as much as that of RMS. The $\Phi_s$ varies from -90° to -86.7° resulting in a longitidunal shrinkage in the bucket. Also, *A*, acceleration effiency gently increases from 0 to 0.004 as a result of small acceleration and there is an increment of 0.30 keV on kinetic energy, *W*, at the end of this section, according to the Figure 2.

Bunching process mainly occurs in GB section in an RFQ. This is carried out by keeping the charge density nearly constant so as to reduce the space charge effects. In this section, parameters, such as $m$, $\Phi_s$, $A$, rise faster than those of other sections because of bunching. In our design, GB consists of 109 cells. At the end of this section, $m$ and $\Phi_s$ have values of 1.90 and -30°, respectively. $A$ is 0.41 whereas $a$ is 2.09 mm in compliance with inverse proportionality to $m$. $W$ is 0.62 MeV as is seen from Figure 2.

In the last section, which is named "Acceleration Section (AS)", there are 83 cells and $m$, $a$ and $\Phi_s$ are nearly constant. Hence, focusing is almost steady for keeping the $A$ high so as to reach desired energy at the end of the RFQ. So, $W$ is 2.99 MeV and $A$ is 0.44 whereas $m$ and $\Phi_s$ are 1.911 and -30° respectively.

Last cell of the RFQ is generally used as a "transition cell" to end the RFQ with quadrupolar symmetry. There is no axial potential, hence, accelerating field in this cell and it makes possible to control the orientation of the ellipse in transverse phase-space. There is also a "fringe field" at the end of this cell that can be used for matching the output beam with the next accelerator structure. A transition cell with a length of 33.98 mm has been used in our design.

As it is seen from Figure 2, the intervane voltage has been chosen as constant along the whole RFQ structure. We have, also, taken the average bore radius and transverse radius of curvature of vane tip, $\rho$, as constant. Thus, capacitance of the vanes is invariant and fabrication becomes easier. This, also, makes a contribution to flatness of, $E_z$, accelerating electric field for the purpose of prevention of particle losses, and to the error analysis without complexity. One of the important factors in determining the intervane voltage is Kilpatrick Criterion [9]. We have chosen a limitation of 1.8 times this criterion considering cw beam has been used. Also, an intervane voltage of 76.80 kV has been kept constant along the RFQ structure. The design parameters obtained from simulation are given in Table 2.

**Table 2.** Design parameters of theRFQ

| Parameters | Value | Unit |
|---|---|---|
| Intervane voltage, $U_0$ | 76.8 | kV |
| Modulation parameter, $m$ | 1 – 1.911 | |
| Average bore radius, $r_0$ | 3.17 | mm |
| $\rho/r_0$ | 0.85 | |
| Synchronous phase, $\Phi_s$ | -90 to -30 | degrees |
| Maximum surface electric field | 31.62 (1.8 Kilpatrick) | MV/m |
| Transmission | 96.9% | |
| Beam power | 86 | kW |
| Power dissipation | 440 (1.7 × SUPERFISH) | kW |
| Total length | 3.45 (without both ends) | m |
| Input emittance (norm., rms), $\varepsilon_{x,y}$ | 0.20 | π·mm·mrad |
| Output emittance (norm., rms), $\varepsilon_x$ | 0.23 | π·mm·mrad |
| $\varepsilon_y$ | 0.23 | π·mm·mrad |
| $\varepsilon_z$ | 0.087 | π·deg·MeV |

As it is seen from Table 2, maximum electric field is 31.62 MV/m in accordance with 1.8 Kilpatrick limit, i.e., 33 MV/m. ~97% of all particles are transmitted and 98.5% of these particles are captured for acceleration. A portion of 86 kW of total RF power requirement of 526 kW is delivered to the beam while the power of 440 kW is dissipated on the structure walls. This value 1.7 times that of SUPERFISH. Also, unaccelerated beam portion of 1.5% has a power of 0.146 kW which is negligible compared to 86 kW. Emittance growth, another figure of merit, is 15% according to the Table 2. Also, the beam has a longitidunal emittance of 0.087 π·deg·MeV at the exit of the RFQ, as an indication of existence of bunching. Brightness of the beam, defined as in Equation (1) [10], is one of the main figures of merit. The beam brightness is 549.5 mA/ $π^2$·mm$^2$·mrad$^2$, referring to Table 2. The output beam profile is shown in Figure 3.

$$B = \frac{I}{\varepsilon_x \varepsilon_y} \qquad (1)$$

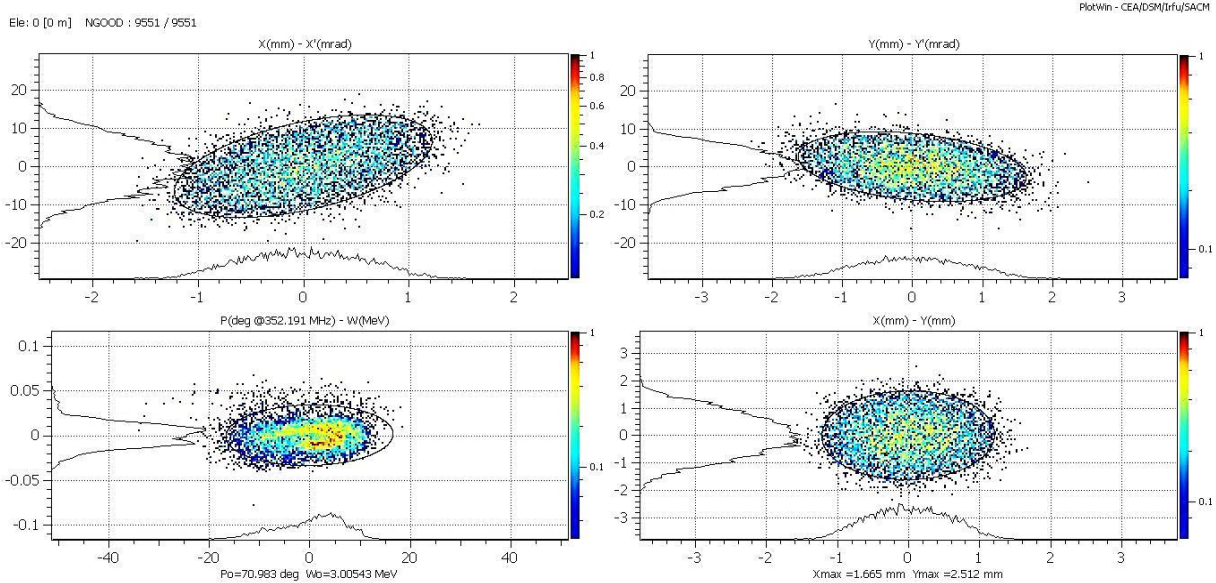

(a)

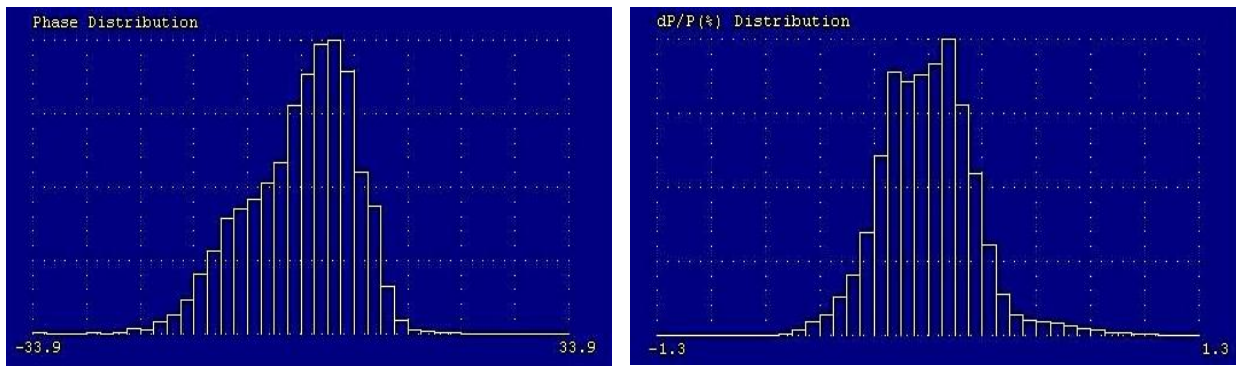

(b)

**Figure 3.** Beam profile (a) and phase and energy distribution of the beam particles (b) at the exit of the RFQ

## 3. Error Study of Beam Parameters

In error analysis of the RFQ, we applied some variations on the input beam parameters to see how the output beam parameters are affected. Transmission and capture (i.e. accelerated particles) efficiencies were figures of merit in this analysis. We have checked over the effects of fluctuations in input beam current, input emittance, input energy and intervane voltage on transmission and capture efficiencies by using LIDOS.RFQ. Simulation results are shown in Figure 4.

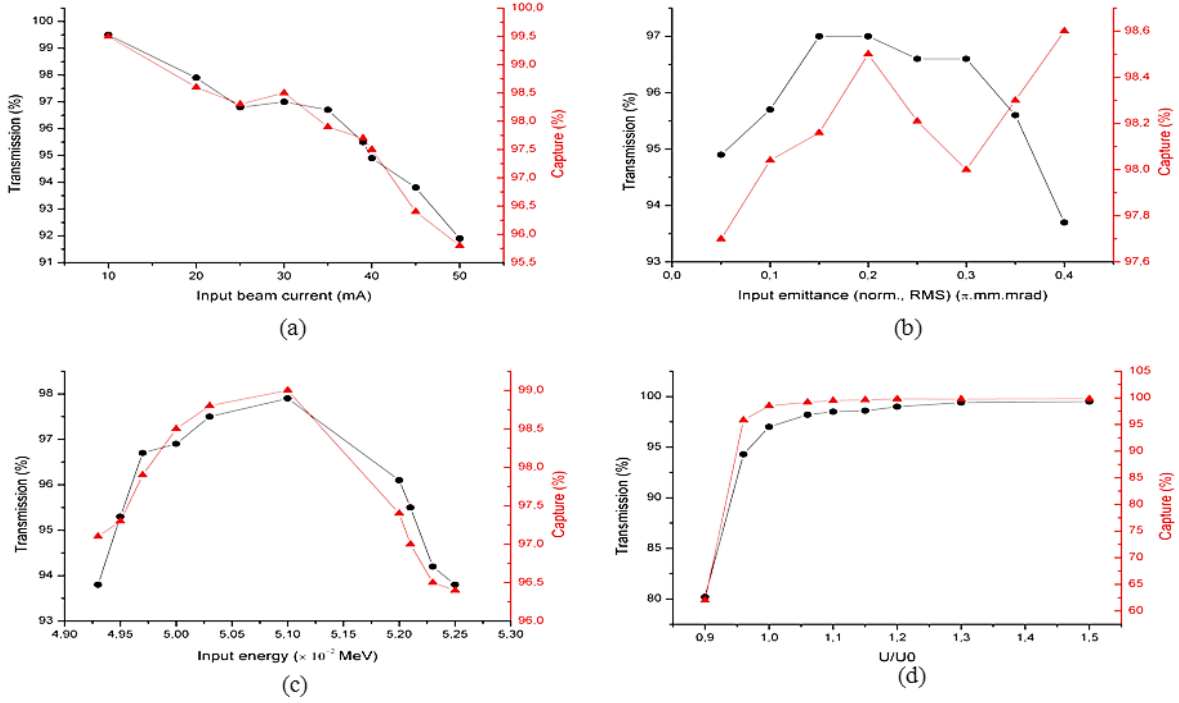

**Figure 4.** Effects of fluctuations in input beam current (a), input emittance (b), input beam energy (c) and intervane voltage (d) on transmission and capture according to the simulation results

In error study simulations, 95% transmission and capture have been chosen as lower acceptability limit. According to this, we can say that an input beam current in the range of 10 mA–40 mA is acceptable, referring to the Figure 4, considering that the other parameters are invariant. Also, input beam energy from 49.5 keV to ~52.2 keV is bearable and input emittance from ~0.07 π·mm·mrad to ~0.37 π·mm·mrad can be tolerated. Besides these, the RFQ works properly if the intervane voltage is started to increase from 75.3 kV ($U/U_0 = 0.98$), although operating voltage is 76.8 kV, paying attention to the Kilpatrick Criterion.

## 4. Two-Dimensional RFQ Cavity Design

The 2-D cross-section of the RFQ cavity design has been done using the computer code SUPERFISH. Various geometrical parameters, describing the RFQ cross-section geometry, have been optimized to attain the rf frequency of 352.2 MHz by the use of this code. This cross-section is the basic element for 3-D models.

RFQfish that is one responsible program in SUPERFISH code group assumes a four-fold symmetry, therefore sets up the geometry for only one quadrant of the RFQ cavity.

All of parameters such as average bore radius ($r_0$), radius of curvature of vane tip ($\rho$), break-out angle ($\alpha_{bk}$) from tip radius to vane-blank width, half width of the blank ($B_w$) have been determined by the beam dynamics simulation and used in SUPERFISH without any manipulation. The gap voltage ($V_g$), the intervane voltage used for the beam dynamics simulation, has been set to 76.8 kV to normalize the electric fields. The $\alpha_{bk}$, used for bit cutting of vane, has been optimized to be 9° and the $B_w$, which must always exceed the $\rho$, has been set to be 7 mm according to the beam dynamics simulation results.

The other geometrical parameters, excepting $r_0$, $\rho$, $B_w$, $\alpha_{bk}$, have been optimized by tuning vane-height ($H$) parameter. Each parameter has been optimized by tuning $H$ as follows: the parameter is optimized once we get the minimum power dissipation and the maximum shunt impedance. The other parameters have been taken as constant during the optimization. For instance, we have optimized $W_s$ by tuning $H$ considering the parameters such as $B_w$, $B_D$, $W_b$, $L_s$, … are constant.

All the optimized parameters obtained from SUPERFISH are listed in Table 3 and full 2-D geometry of the RFQ electrodes constructed by CST MWS regarding to SUPERFISH results in Figure 5. Electric field pattern is seen in Figure 6 in which the details around the vane tips are also shown.

**Table 3**. Two-dimensional design parameters of RFQ

| Parameters | Value | Unit |
|---|---|---|
| **Resonant Frequency** | 352.16 | MHz |
| **Adjacent Dipole Mode Frequency** | 341.42 | MHz |
| **Quality Factor, $Q$** | 11145.3 | |
| **Average Bore Radius, $r_0$** | 0.317 | cm |
| **Transverse Radius of Curvature, $\rho$** | 0.269 | cm |
| **Break-Out Angle, $\alpha_{bk}$** | 9 | degrees |
| **Vane-Blank Half Width, $B_w$** | 0.7 | cm |
| **Vane-Blank Depth, $B_D$** | 3.6 | cm |
| **Vane Shoulder Half Width, $W_s$** | 0.705 | cm |
| **Vane Base Half Width, $W_b$** | 1.4 | cm |
| **Vane Shoulder Length, $L_s$** | 1.712 | cm |
| **Vane Height, $H$** | 9 | cm |
| **Vane Half Width, $W$** | 4.177 | cm |
| **Corner Radius, $R_c$** | 1.678 | cm |
| **Vane Angle 1, $\alpha_1$** | 12.5 | degrees |
| **Vane Angle 2, $\alpha_2$** | 19.5 | degrees |

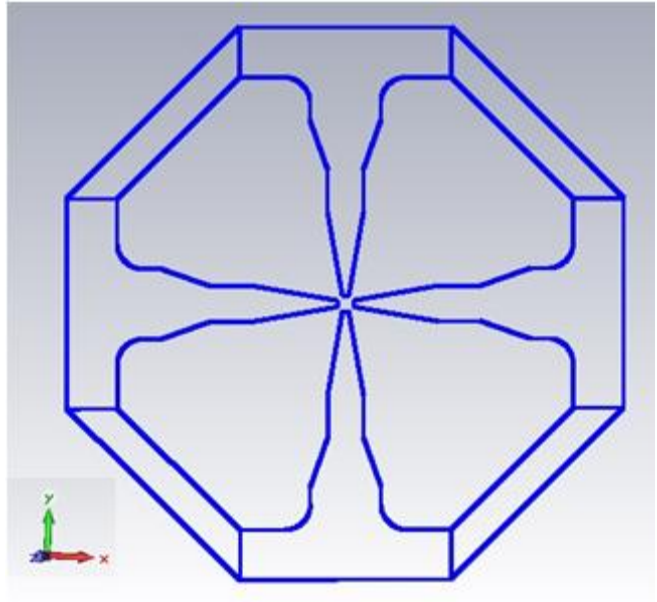

**Figure 5.** A full 2-D geometry of the RFQ electrodes constructed by CST MWS according to SUPERFISH results. The beam flows inward on *z*-axis

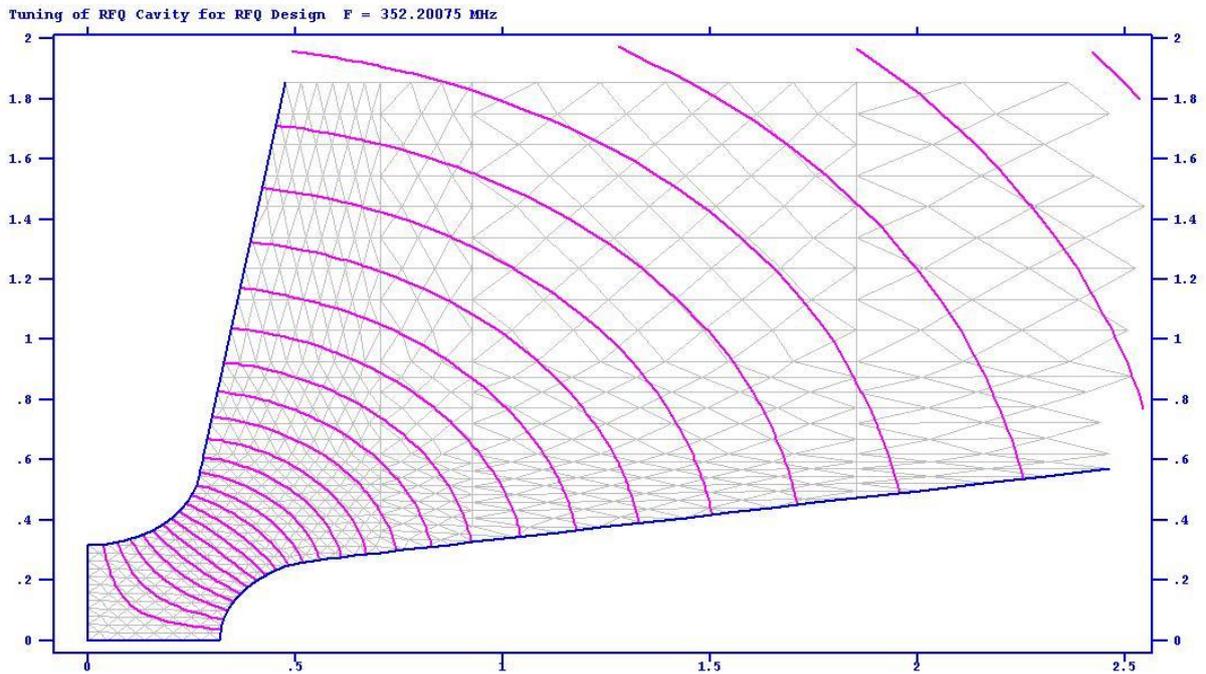

**Figure 6.** 2-D geometry of one quadrant of the RFQ electrodes obtained from SUPERFISH. The electric field pattern between the electrodes is indicated by lines.

Although RFQ works at quadrupole mode frequency, the frequencies of undesirable modes, such as dipole mode, can distort the quadrupole mode and cause unstabilities. Thus, quadrupole ($TE_{210}$-like) and dipole ($TE_{110}$-like) modes frequencies have been calculated applying appropriate boundary conditions, once the parameters had been optimized for one quadrant of the RFQ. The quadrupole mode frequency has been determined as 352.16 MHz by applying Neumann boundary condition around the vane-shoulder half width ($W_s$) in Figure 5 and Dirichlet boundary condition around the vane-tips (Figure 5) while the dipole mode frequency has been obtained as 341.42 MHz changing the boundary condition around the vane-tips to Neumann boundary. The difference of ~11 MHz between two modes is sufficient enough. This difference depends on the RFQ length because the longer the RFQ, the closer the high-order modes come to the operating mode [11]. The quality factor for quadrupole mode has been calculated to be 11145.3 using SUPERFISH.

## 5. Three-Dimensional RFQ Cavity Design

Looking into the more details of electromagnetic field properties in the complex structure of the RFQ cavity and benchmarking 2-D model of RFQ cavity are possible with CST Microwave Studio because of its large mesh ratio. CST MWS also takes the advantage of Perfect Boundary Approximation (PBA) technique which delivers fast convergence in short time [12].

Firstly, we have prepared the full 2-D RFQ model using the geometrical parameters got from SUPERFISH. Later, this 2-D model has been extended to unmodulated 3-D RFQ cavity model as is seen from Figure 7.

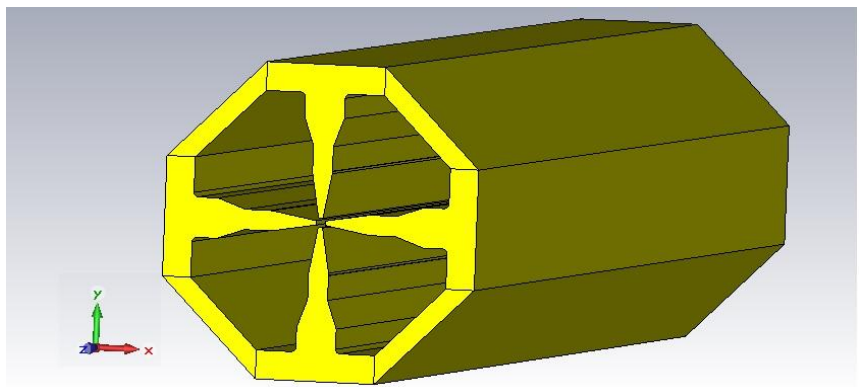

**Figure 7.** The three-dimensional unmodulated RFQ model

The right boundary conditions should be applied to simulate the correct resonant frequency of the structure. In the *x* and *y*-directions, the boundaries are electric ($E_t=0$) while the boundaries are magnetic ($H_t=0$) in the *z*-direction. For the quadrupole mode, the magnetic boundary conditions ($H_t=0$) are put at the both *xz* and *yz*-planes. But the boundary conditions at *xz* and *yz*-planes should not be same for the dipole mode [13]. The quadrupole mode and dipole mode frequencies obtained from CST MWS are 352.17 MHz and 345.37 MHz, respectively. These values are close to those of SUPERFISH.

## 6.  Conclusions

A conceptual design of 352.2 MHz and 3 MeV RFQ for TAC linear proton accelerator has been performed out of deference to beam dynamics. A 4-D uniform beam with 30 mA current and 50 keV kinetic energy has been used in simulations done by using LIDOS.RFQ software. These current and energy values have been chosen in accord with the latest feasibility studies. Minimum emittance growth, compactness of the RFQ structure and beam transmission were figures of merit during beam dynamics simulation. Some beam dynamics parameters such as *m* and $\Phi_s$, have been optimized in the existence of space-charge effects. A transmission of ~97% with 98.5% capture and an emittance growth of 15% have been obtained after the optimization. Also the optimized RFQ has a length of 3.45 m without the end caps on both sides. Such an RFQ requires an total RF power of 526 kW according to the simulation results. Error analysis has also been done in this study, introducing some variations in the input beam parameters, to see how the output beam parameters are affected. Tolerance limits belonging to input beam current, input beam emittance, input beam energy and intervane voltage have been specified in this analysis.

The 2-D electromagnetic structure design of the RFQ has been done by using SUPERFISH code to prevent the distortions on quadrupole mode frequency induced by the nearest dipole mode. According to SUPERFISH results, the difference between these two modes is 11 MHz, roughly. 2-D electromagnetic design has also given a high quality factor, *Q*, of 11145.

The 3-D electromagnetic structure design has been done for more detailed electromagnetic structure view of the RFQ by using CST MWS software since it has a large mesh ratio. The quadrupole mode frequency and *Q* are 352.17 MHz and 11677, respectively, while the dipole mode frequency is 345.37 MHz which are compatible with SUPERFISH. Based on these results, the TAC RFQ has good parameters and is less sensetive to small variations in input beam parameters. The next work belonging to the RFQ would be detailed RF analysis including thermal analysis, accelerating electric field stabilization (via cut-backs of end-vanes, and π-mode stabilizers, if it is needed [14]) which is another important and challenging step for the RFQ design.

## 7. Acknowledgements

We thank Turkish State Planning Organization (DPT), under the grants no DPT-2006K120470 for their supports and the Turkish Atomic Energy Authority (TAEK) SNRTC for benefiting from their computing facilities. We also special thank Alessandra Lombardi from Linac4/CERN and Saleh Sultansoy from TOBB University of Economics and Technology for their helps.

## REFERENCES

[1] Sultansoy, S., "*Regional project for elementary particle physics: linac-ring type c-τ-factory*", Turk. J. Phys. 17, 591-597 (1993).

[2] Turkish Accelerator Center web page: http://thm.ankara.edu.tr

[3] Arik, M., Bilgin, P.S., Caliskan, A., et al., "*A Provisional Study of ADS within Turkic Accelerator Complex Project*", III. International Conference on Nuclear&Renewable Energy Resources (NuRER 2012), Istanbul, TURKEY, 20-23 May 2012.

[4] Staples, J.W., "*RFQ's- An Introduction*", AIP Conference Proceedings 249, New York, 1992, p 1483.

[5] G.H. Gillespie Associates, Inc. and The LIDOS Group, "*LIDOS.RFQ.DESIGNER Version 1.5 User's Guide*", Accelsoft Inc., USA.

[6] Billen, J.H. and Young L.M., "*Poisson Superfish*", LA-UR-96-1834.

[7] CST Studio Suite, "*CST Microwave Studio*", 2008. http://www.cst.com

[8] Wangler, T.P., "*Principles of RF Linear Accelerators*", John Wiley&Sons, New York, (1998), p 178.

[9] Kilpatrick, W.D., "*Criterion for Vacuum Sparking Designed to Include Both rf and dc*", The Review of Scientific Instruments, Volume 28, number 10, 1957.

[10] Buon, J., "*Beam Phase Space and Emittance*", CAS, General Accelerator Physics Course, Jülich, 17-28 September 1990 and LAL/RT 90-15.

[11] Vretenar, M., "*The Radio Frequency Quadrupole*", CERN Accelerator School (CAS) High Power Hadron Machines, Bilbao, 2011.

[12] Web page of Computer Simulation Technology: https://www.cst.com/Products/CSTmws/FIT, (22.05.2014).

[13] Li, D., Staples, J.W. and Virostek S.P., "*Detailed Modeling of the SNS RFQ Structure with CST Microwave Studio*", Proceedings of LINAC 2006, Knoxville, Tennessee, USA, 21-25 August, 2006, pp 580-582.

[14] Rossi, C., Bourquin, P., Lallement, J.-B., et al. *"The Radiofrequency Quadrupole Accelerator for the Linac4"*, Proceedings of LINAC08, Victoria, BC, Canada, 29 September-03 October 2008, pp 157-159.